\documentclass[letterpaper, 10 pt, conference]{ieeeconf}  

\IEEEoverridecommandlockouts                              

\usepackage{graphicx}
\usepackage{algorithmic}
\usepackage{algorithm2e}

\title{\LARGE \bf Cybersecurity Threats in Connected and Automated Vehicles based Federated Learning Systems}

\author{Ranwa Al Mallah$^{1}$, Godwin Badu-Marfo$^{2}$, Bilal Farooq$^{3}$\\ \\
Forthcoming in Proceedings of IV21
\thanks{* This work is supported by NSERC Canada Research Chair Program}
\thanks{$^{1}$Ranwa Al Mallah is with the Laboratory of Innovations in Transportation,
        Ryerson University, ON, Canada
        {\tt\small ranwa.almallah@ryerson.ca}}%
\thanks{$^{2}$Godwin Badu-Marfo is with the Laboratory of Innovations in Transportation,
        Ryerson University, ON, Canada
        {\tt\small @ryerson.ca}}%
\thanks{$^{3}$Bilal Farooq is with the Laboratory of Innovations in Transportation,
        Ryerson University, ON, Canada
        {\tt\small bilal.farooq@ryerson.ca}}%
}

\begin{document}

\maketitle
\thispagestyle{empty}
\pagestyle{empty}

\begin{abstract}

Federated learning (FL) is a machine learning technique that aims at training an algorithm across decentralized entities holding their local data private. Wireless mobile networks allow users to communicate with other fixed or mobile users. The road traffic network represents an infrastructure-based configuration of a wireless mobile network where the Connected and Automated Vehicles (CAV) represent the communicating entities. Applying FL in a wireless mobile network setting gives rise to a new threat in the mobile environment that is very different from the traditional fixed networks. The threat is due to the intrinsic characteristics of the wireless medium and is caused by the characteristics of the vehicular networks such as high node-mobility and rapidly changing topology. Most cyber defense techniques depend on highly reliable and connected networks. This paper explores falsified information attacks, which target the FL process that is ongoing at the RSU. We identified a number of attack strategies conducted by the malicious CAVs to disrupt the training of the global model in vehicular networks. We show that the attacks were able to increase the convergence time and decrease the accuracy of the model. We demonstrate that our attacks bypass FL defense strategies in their primary form and highlight the need for novel poisoning resilience defense mechanisms in the wireless mobile setting of the future road networks. 

\end{abstract}

\section{INTRODUCTION}

Federated Learning (FL) is a machine learning technique that allows to address issues such as data privacy, security and access rights. However, despite the advantages, there are still critical challenges in applying FL to emerging wireless mobile networks e.g. Connected and Automated  Vehicles (CAV) based Intelligent Transportation Systems (ITS) \cite{niknam2020federated}. 


The road traffic network represents an infrastructure-based configuration of a mobile and wireless network on which CAVs travel, use regulated frequencies, and have access to the bandwidth to communicate. Unlike mobile phones that communicate through a high-speed network, CAVs exchange V2X messages with unknown moving vehicles, Road Side Units (RSU), pedestrians, and cyclists with no prior association. Vehicle-to-Vehicle (V2V) messages enable vehicles to exchange information with other surrounding vehicles in order to prevent incidents or traffic conditions. Vehicle-to-Infrastructure (V2I) complements V2V communications and enable RSUs to exchange information with the vehicle. These technologies exchange packets called Basic Safety Messages (BSM) designed to contain no personal identifiable information, since anonymity of sender must always be maintained. Vehicles and their drivers should remain untraceable in order to ensure the privacy in ITS. 

Recent reports identified highly practical wireless attacks on CAVs \cite{dibaei2019overview}. Some attacks target in-vehicle security and others target security of inter-vehicle communications \cite{mothukuri2021survey}. Currently, there is no security mechanism in place to validate and authenticate basic safety messages and ensure trusted communication among the random moving entities. 
A corrupted device in the vehicle can result in false BSM exchanged even though the sender is trusted \cite{williams2017danger}. 



Vehicular networks are highly heterogeneous because vehicles may have a range of different communication technologies. Unfortunately, a direct application of existing FL protocols without any consideration of the underlying communication infrastructure of the CAVs will expose the FL process to cyberattacks. For instance, malicious entities may exploit vulnerabilities in the vehicular network in order to poison the training of the model with false inputs. The existing defense algorithms are more suitable to cloud assisted applications or data centers. 


Recent work has started to address the classical problems in FL in the context of vehicular networks, such as privacy, large-scale machine learning, and distributed optimization \cite{li2020federated}. However, research on the cybersecurity issues related to FL in the context of vehicular networks has yet to be explored. In fact, in terms of security, similar to machine learning approaches in fixed networks, an attacker can data-poison the FL process or can perform model poisoning. Particularly, model poisoning attacks affect the convergence of the global model because of the malicious local model updates that the attacker may send back to \textit{chief} node of the FL process. In \cite{bhagoji2019analyzing}, FL has been analyzed through an adversarial lens to examine the vulnerability of the learning process to the model-poisoning adversaries in a fixed configuration network setting.

In this paper, we aim at analyzing the vulnerability of the implementation of FL in mobile wireless vehicular networks. We explore how the distributed training gives rise to a new type of threat in the mobile environment that is different from traditional fixed networks. This threat is due to the intrinsic characteristics of the wireless medium and is caused by the highly mobile nature of CAVs. We explore falsified information attacks which target the FL process that is ongoing at the RSU. We highlight a number of attack strategies on FL in this context. In fact, an attacker can compromise a vehicle and perform falsified information attacks. By continuously driving through the same street and performing a model poisoning attack, it can overcome the effects of other CAVs and disrupt the training of the global model. We then design an attack where a single CAV is able to declare multiple identities and perform a critical poisoning via model replacement at convergence time. Because of the vulnerabilities in the environment of the FL process, our proposed attacks bypass classic defense strategies in their primary form. Already proposed defense techniques do not seem to fit the dynamic and random unknown CAV ecosystem, where vehicles are spread and move across a geographical area. Our attacks highlight the need that novel poisoning resilience defense mechanisms are urgently required in the mobile wireless settings.

This work is organized as follows: Section 2 presents a brief literature review. The FL protocol in a wireless mobile setting is presented in Section 3. The threat model and the attacks are illustrated in Section 4. In Section 5 we present the simulation environment and the results. Finally, concluding remarks and future outlook are in Section 6.

\section{LITERATURE REVIEW}

Wi-Fi, WiMAX, Long-Term Evolution (LTE), Near-Field Communication (NFC), and Dedicated Short-Range Communications (DSRC) are among the communication technologies available for vehicular data communications. In reviewing the literature, several attacks on the communication network involving ITS were found \cite{dibaei2019overview}. 

In FL, the adversarial goal of model poisoning attacks is to ensure that the global model converges to `sub-optimal to utterly ineffective models,' while the defense aims at ensuring the convergence \cite{mhamdi2018hidden}. Among the defense strategies, secure aggregation mechanisms are proposed for the distributed learning to ensure convergence \cite{blanchard2017machine}. Other defense mechanisms used clustering to detect model updates that are different to what they should be. Rather than correcting, more recently, some solutions proposed measures designed to detect malicious \textit{workers} by interpreting information of the worker's behavior \cite{kang2020reliable}. However, these defense mechanisms are more realistic in a data center setting, where the entities are dedicated to a server. The server in this setting is able to identify and authenticate the entities of its underlying network. In contrast, in wireless mobile federated networks and with the coexistence of dedicated short-range communication and cellular-connected vehicle-to-everything (c-V2X) in the same ITS band, each vehicle is free to join and leave the network and most vehicles are not active on any given iteration. Thus the FL vehicle training poses multiple novel challenges. To the best of our knowledge, no previous work has studied the effects of this more realistic node-centric communication scheme in which each vehicle can decide when to interact with the road side unit in an event-triggered manner. This paper exposes the key vulnerabilities of FL implementations in wireless mobile networks by proposing attacks that exploit the mobility of the connected and automated vehicles. The threat comes from CAVs exploiting the medium to perform model poisoning attacks on the FL model. 


\section{FL PROTOCOL IN A WIRELESS SETTING}

\begin{figure}
    \centering
    \includegraphics[width=1\columnwidth]{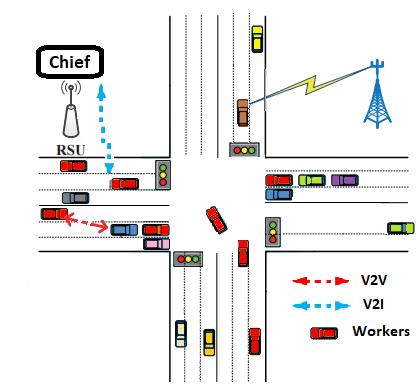}
    \caption{Vehicles acting as \textit{workers} in the FL process communicate using V2I with the \textit{chief}
implemented at the Road Side Unit.}
    \label{fig:FL_fig}
\end{figure}

We consider a vehicular network where a roadside unit can take the role of the \textit{chief} and the vehicles act as \textit{workers}, as can be seen in Figure~\ref{fig:FL_fig}.

However, in its current form, the FL protocol does not directly apply to vehicular networks. Bonawitz et al. described in details the FL protocol in a wireless mobile environment where the \textit{workers} are mobile phones, \cite{bonawitz2019towards}. They considered 'eligible for training' phones that are on idle, charging, and connected to an unmetered network such as WiFi. This is problematic for vehicular applications and cannot be expected from CAVs who are envisioned to participate in the FL process while navigating on the road network. Most vehicular applications require real-time microscopic and macroscopic traffic variables to be extracted by the vehicles and it is the dynamic with the surrounding vehicles that will lead to superior modeling at the application level. Offline training may introduce bias in the training and may lead to inferior models. Therefore, since the protocol proposed by Bonawitz et al. is not particularly configuration specific, we adapt it for the vehicular network in our study.

\noindent
\textbf{FL Protocol for vehicular network:}

\begin{itemize}

\item The RSU identifies an application and its learning problem and broadcasts the FL task to the vehicles in its coverage area. An FL task is a specific computation such as training to be performed with given hyperparameters e.g. learning rate, batch size and number of epochs to run.
 

\item Vehicles must stay connected to the \textit{chief} for the duration of the round. The \textit{chief} can either consider all the vehicles announcing their availability as \textit{workers} or select a subset to work on the FL task. 


\item Once a round is established, the \textit{chief} then broadcasts the current global model parameters.

\item Each selected \textit{worker} then performs a local computation based on the global model and its local dataset, and sends a local model update back to the \textit{chief}. A notable advantage of FL in this setting is that it does not rely on synchronization among the \textit{workers}. Hence, even during a loss of connectivity between the vehicles and the RSU, vehicles can still build their local models and navigate; this is crucial in a very dynamic environment as long as the \textit{worker} adhere to the time window in the timing plan.

\item As local model updates are received, the \textit{chief} aggregates them using federated averaging. If enough \textit{workers} report in time, the round will be successfully completed and the \textit{chief} will update its global model, otherwise, the round is abandoned. The \textit{chief} incorporates the updates into its global model, and the process repeats.

\end{itemize}

The integrity of the local model updates sent by the \textit{workers} play a crucial role on the performance of the collaborative learning model. In fact, the overall training process can be compromised by some \textit{workers} intentionally providing wrong local model update values to the RSU in order to significantly affect convergence time. Current defense schemes are not quite suitable for vehicular networks because they do not consider the threats coming from the mobility of the CAV.

\section{THREAT MODEL}

The reliability of the applications of ITS is highly dependent on the quality of the data collected across the traffic network. In this paper, we describe a yet unexplored threat model that targets data integrity in federated learning.

In a normal FL setting, vehicles are able to continuously extract link level information and use them locally to participate, in a distributed manner, in the training of a global model. Vehicles travel the zone under the coverage of the RSU and then broadcast their local model update to the RSU who is responsible of the aggregation in the FL process. RSUs are a core component in vehicular networks as they must authenticate, manage and update users and their transmitted messages. Therefore, a successful attack on a RSU can have a detrimental effect on its operations. This paper explores falsified information attacks, which target the FL process that is ongoing at the RSU. 

\textbf{Attack1. Standard falsified information attack} 

In the falsified information attack, compromised information is sent out by a malicious vehicle that is moving in and out of the zone under study very rapidly and thus continuously providing falsified real-time updates to the RSU. The zone under study represents the area where the RSU can receive messages. In this scenario, a single attacker designs malicious local model updates and sends them to the RSU to target the training of the model that is ongoing at the RSU. The adversary's aim is to prevent convergence of the global model. This attack, if effective, represents a strategy that allows for maximum results with minimum effort due to the fact that it is not computationally expensive to execute. Also, this attack is hard to mitigate because the incoming messages may come from a legitimate vehicle having a valid, authenticated on board unit and having credentials to be considered an authorised vehicle. Moreover, a vehicle can impersonate a legitimate vehicle and launch falsified information attacks. An impersonation attack is the result of a man-in-the-middle attack where a vehicle intercepts messages exchanged through the traffic network looking for one with authentication information. Once the attacker has access to credentials of authorised vehicles, he uses them to impersonate authorised vehicles and launch the attacks. Although a number of vehicular ad-hoc networks robustness approaches currently in the literature would be effective in preventing impersonation attacks, no defenses are specifically geared towards situations in which this underlying assumption holds. Moreover, since impersonation is not the only assumption of our attack model, as of today, a legitimate vehicle can attack without being detected although defense techniques against impersonation would not enable an attacker to use stolen credentials. Algorithm~\ref{alg:Attack_simple} is implemented at the CAV and aims at conducting an untargeted model poisoning attack on a federated learning task that is ongoing at the RSU. 

\begin{algorithm}

\SetKwInput{KwInput}{Input} 
\SetKwInput{KwOutput}{Output}
\SetKwFunction{FMain}{Main}
\KwInput{Set of road segments under the coverage of the RSU, \textit{$RS_c$}, Global Model, $GM^t$, parameters sent by the chief sent at iteration \textit{t} to the \textit{workers} in its coverage area.}
\KwOutput{Poisoned local model updates $LM_{i,t}$ sent by the CAV at iteration \textit{t}.}
\SetKwProg{Fn}{Function}{:}{}
\Fn{\FMain{}}{
\If {Current segment is in \textit{$RS_c$ }}{
   Get $GM^t$ parameters\;
   Reply to the \textit{chief} to participate in training\;
   \If {Instructions and timing plan received}{
   Create $LM_{i,t+1}$ with random parameters sampled from a distribution\;
   Send $LM_{i,t+1}$ to the \textit{chief}\;
   }
   }
   }
 \caption{Algorithm of the standard falsified information attack conducted by a single CAV on a federated learning process.}
 \label{alg:Attack_simple}
\end{algorithm}

\textbf{Attack2. Sybil attack} 

The Sybil attack can be seen as a variant of the falsified information attack, an evolved version of it. A Sybil attack consists in one vehicle creating fake vehicle identities and using them to broadcast local model updates that may compromise the FL process. In this scenario, the vehicle transmits multiple messages each with a different ID. The IDs could have been spoofed or stolen from compromised vehicles. This will enable the attacker to fabricate false messages and have a greater influence on the FL process. Each round, the FL protocol randomly selects vehicles to participate in the training, a Sybil attack would allow the attacker to increase it chances to be selected in the process. The increase in the number of malicious vehicles will potentially impact the training and shift the global model away from convergence. The attacker may perform a critical attack via model replacement at convergence time by simultaneously sending falsified local model updates. This strategy makes it difficult for mitigation methods because it is almost impossible to predict a potential malicious behavior as it happens suddenly by the group of Sybils. Algorithm~\ref{alg:Attack_sybil} is implemented at the CAV and aims at conducting a sybil attack on a federated learning task that is ongoing at the RSU. 

\begin{algorithm}

\SetKwInput{KwInput}{Input} 
\SetKwInput{KwOutput}{Output}
\SetKwFunction{FMain}{Main}
\KwInput{Set of road segments under the coverage of the RSU, \textit{$RS_c$}, Global Model $GM^t$ parameters sent by the chief sent at iteration \textit{t} to the \textit{workers} in its coverage area.}
\KwOutput{Poisoned local models updates $LM_j{(i,t)}$ sent by the CAV at iteration \textit{t}.}
\SetKwProg{Fn}{Function}{:}{}
\Fn{\FMain{}}{
\If {Current segment is in \textit{$RS_c$ }}{
   Get $GM^t$ parameters\;
   Generate set of Sybil nodes, \textit{Syb}\;
   \For{\textit{j} in \textit{Syb}}{
   Reply to the \textit{chief} to participate in training\;
   \If {Instructions and timing plan received}{
   Create $LM_j{(i,t)}$ with random parameters sampled from a distribution\;
   Send $LM_j{i,t+1}$ to the \textit{chief}\;
   }
   }
   }
   }
 \caption{Algorithm of the Sybil attack conducted by the CAV on a federated learning process.}
 \label{alg:Attack_sybil}
\end{algorithm}


\indent Falsified information attacks can substantially degrade the performance of the FL process that rely on averaging to generate the global model. In Attack1, a persistent attacker will have an impact, however in the Sybil attack, attackers aim at producing a larger attack impact and by the same time avoid detection. By creating fake vehicle identities, the attacker is not only motivated by the greater scale of the attack, but also seeking to hide it from intrusion detection systems that may raise alarms if the attack is originating from a single vehicle identity. It is worth mentioning that in the case of a Sybil attack, the number of fake identities generated onto a single traffic lane should adhere to the traffic flow capacity of the lane to avoid raising intrusion flags. 



\section{SIMULATION AND RESULTS}

We evaluate our attack strategies on the predictive model for link level speed, developed using a deep learning based time series model. Predicting speed on a road segment enables traffic managers to take early actions to control flow and prevent congestion. In \cite{alfaseeh2020deep}, using the technological advancements related to CAVs, they trained a Long Short Term Memory (LSTM) deep network to predict speed on a link. In fact, the estimation of the parameters of the LSTM using data samples gathered at each vehicle is crucial for the prediction of link average speed. The authors modeled the average speed distribution using a central entity, which is the RSU to compute and communicate with all the vehicles at each time step. However, this centralized approach may be impractical due to: (i) The excessive overhead needed to communicate with all the vehicles in the dynamic wireless mobile network will degrade the network-wide performance, and (ii) vehicles wanting to keep their data private by not sharing it with other vehicles, in which warrants collaborative learning methods. Therefore, we propose the implementation of a distributed method for the predictive modeling of speed in a practical vehicular network. The solution is based on FL to train an LSTM model by allowing each vehicle to learn a local model individually using local observations that are never communicated to the RSU.

\subsection{Simulation outline}	

Our experiments utilize downtown Toronto’s road network as it experiences high levels of congestion, specifically during the morning peak period. The road network covers 76 intersections and 223 links. The vehicular demand is provided by the Transportation Tomorrow Survey (TTS) for the 7:45am and 8:00am peak period for the year 2014. To extract realistic measurements at every second, we implemented the scenarios in Matlab with two layers, the communication and the physical layer representing the microscopic traffic simulation as in  \cite{djavadian2018distributed}. In fact, the agent based simulator used in this study implemented a calibrated Intelligent Driver Model (IDM) for vehicular movement. It is the car-following adopted for the displacement estimation at every second, which is used to calculate the speed of vehicles. The second-by-second vehicular characteristics are captured and then used to estimate the space mean link indicators. Simulation ends once all of the vehicles reach their destinations. Vehicular characteristics are captured and used to estimate space mean link indicators. 

We aim at training an LSTM network to predict the average link speed. The LSTM network consists of five hidden layers in a setting of three sequences of speed, density, and in-links speed. We tuned several hyper-parameters such as the learning rate, epochs, learning rate drop factor, momentum and the number of hidden units of the different layers. The FL protocol then iteratively demands vehicles to download this model from the RSU, update it with their local observations, and upload the updated model to the RSU. The RSU then aggregates the multiple model updates received by the vehicles to generate a global model in order to then repeat the process.

\subsection{Results}	

\begin{figure}
    \centering
    \includegraphics[width=1\columnwidth]{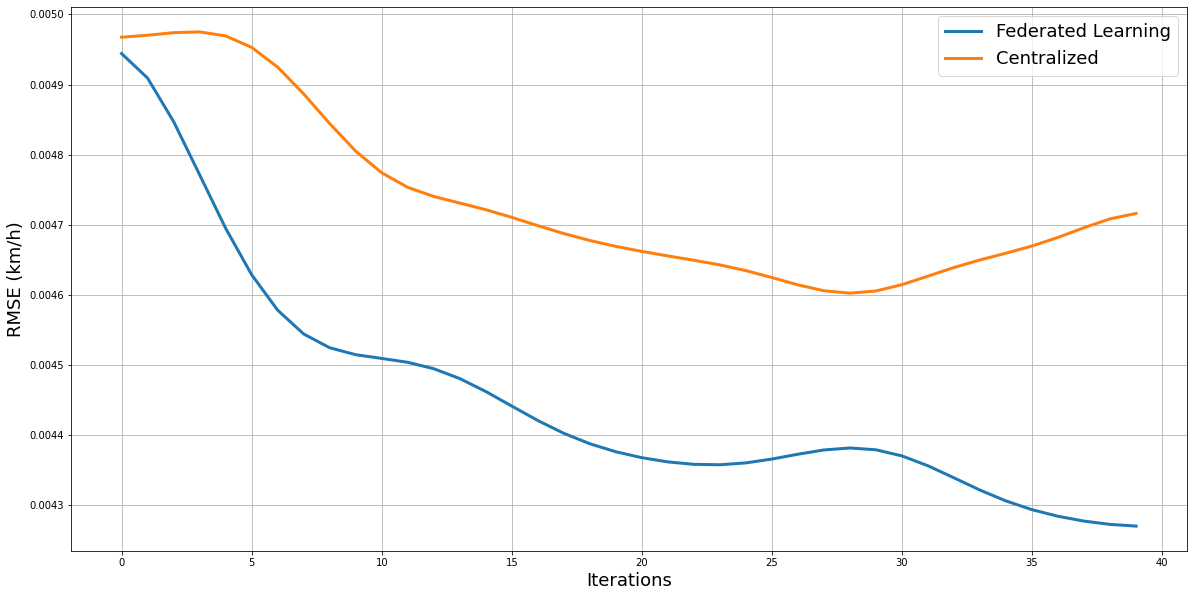}
    \caption{Prediction accuracy of the LSTM model in a centralized and federated learning setting (baseline).}
    \label{fig:Prediction_centralizedML}
\end{figure}

We present in Figure~\ref{fig:Prediction_centralizedML} the results for the centralized training of the LSTM at the RSU for the prediction of average speed. In this context, all vehicles upload via V2I their data samples to the RSU which estimates the global model. We show in the figure the prediction performance of the model in terms of Root Mean Square Error (RMSE) under the centralized setting. The RMSE value of the speed predictive model is tending to approximately 0.00475 km/h.

We then implement the FL protocol proposed in this paper for the vehicular application under study. Current FL algorithms such as federated averaging can efficiently utilize hundreds of devices in parallel even though many more may be available. In the vehicular network of our study, and specifically for the use case application considered, the topology of the network is dictating the maximum number of \textit{workers} to be considered by the algorithm and does not require increased parallelism. In fact, the maximum link length is approximately 450 meters. The speed range is from 0 to 80 km/h. Under the free flow traffic condition, the maximum travel time required to traverse a link is around 0.8 minute. This bounds the number of \textit{workers} to be considered with an upper bound corresponding to the maximum density on the link. We fixed the number of \textit{workers} $\mathcal{K} = 10$. In low density, when the number of vehicles travelling the road segment is $\mathcal{K} \leq 10$, all the agents are chosen at every round of training to ensure that a sufficient number of vehicles connect to the \textit{chief}. Otherwise, 10 \textit{workers} are chosen at random every iteration. This is important both for the rate of task progress at the \textit{chief} and for the security properties of the secure aggregation protocol. Also, it avoids excessive activity during peak hours without hurting FL performance. Moreover, our assumption holds because in a real world deployment, at any point in time only a subset of vehicles connect to the \textit{chief} due to disconnections and drop out due to computation errors, network failures, or changes in eligibility is high in this setting. It is worth noting that the number of participating vehicles depends on the time of day and the road traffic condition. We run federated learning until a pre-specified test accuracy is reached or the maximum number of time epochs have elapsed. 

We observe in Figure~\ref{fig:Prediction_centralizedML} that FL enables the proposed distributed method to estimate the average speeds on the segment with an accuracy that is comparable to that of the centralized solution. This result was expected since depending on the federated optimization algorithm used, federated has comparable performance to centralized learning when using i.i.d. data, but not with non-i.i.d. data \cite{nilsson2018performance}. 

\begin{figure}
    \centering
    \includegraphics[width=1\columnwidth]{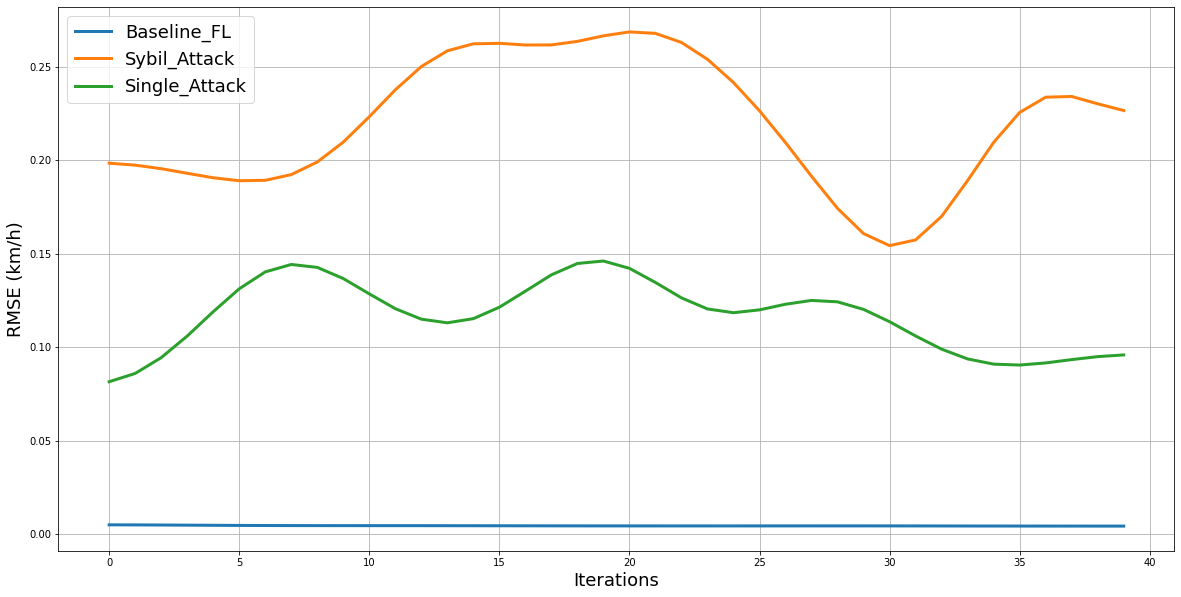}
    \caption{Prediction accuracy of the LSTM model under different types of attacks in a federated learning setting.}
    \label{fig:Prediction_FL}
\end{figure}

In Figure~\ref{fig:Prediction_FL}, we demonstrate the vulnerability of federated learning to our attacks. From a single corrupted vehicle, an attacker can sabotage the system and we show how different attack modulations and parametrisations affect the training. Firstly, we show the impact of the falsified information attack conducted by the CAV on the FL process. We notice that in terms of prediction accuracy, the RMSE value increased of approximately 0.16 km/h for the Single\textunderscore Attack compared to the baseline, Baseline\textunderscore FL that is under no attack. In this scenario, although acting alone, the attacker was able to gradually increase the error on the prediction by continuously traversing the link and causing poisoning of the global model.

Then, we present on the same Figure~\ref{fig:Prediction_FL} the impact of the Sybil attack. We proposed the Sybil attack to improve stealth and have the same CAV cause a bigger impact on the FL process. In the figure, we notice the increase in RMSE by 0.15 compared to the baseline and 0.04 compared to Single\textunderscore Attack. By fabricating fake vehicle identities, the CAV was able to conduct a single shot attack and resulted in an increase in the error. 

Defense strategies aiming at correcting the aggregation rule cannot detect the malicious model updates being continuously sent by the CAVs in both the Single\textunderscore Attack and the Sybil\textunderscore Attack because primarily it is not the model update that is malicious but the behavior of continuously sending irrelevant models that is malicious. The models sent by the attacker are fabricated to delay convergence because they do not contribute to the learning by actively training a local dataset. Also, the CAV is allowed to traverse the road segment as much as he desires as long as he respects the speed limit. 

\begin{figure}
    \centering
    \includegraphics[width=1\columnwidth]{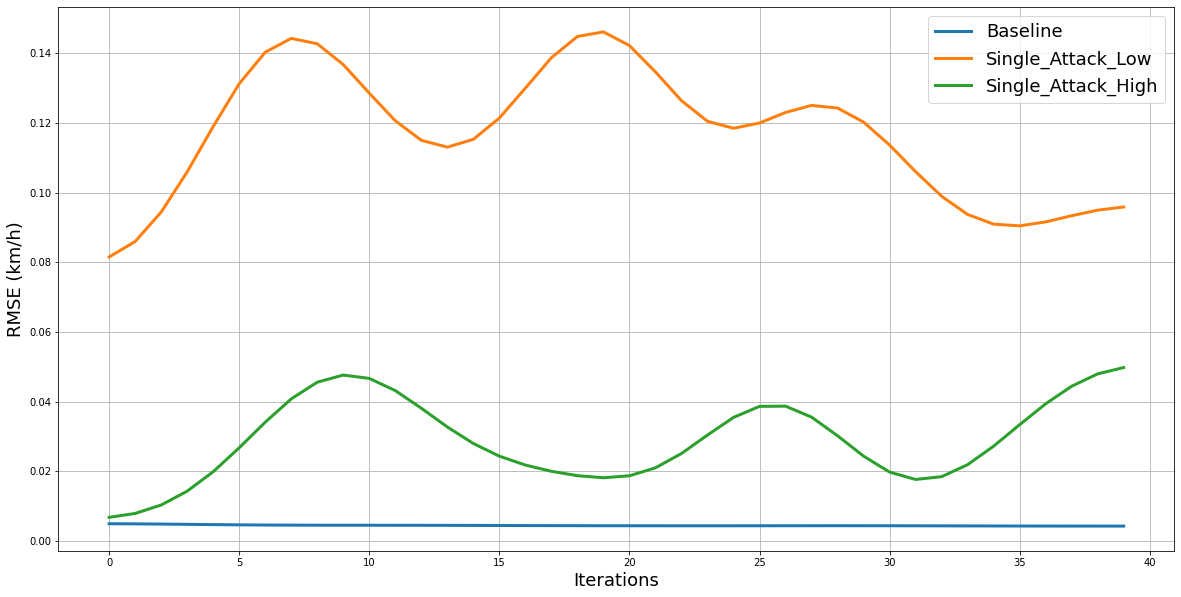}
    \caption{Impact of the Single\textunderscore Attack in different traffic densities.}
    \label{fig:Prediction_FL_Densities_1}
\end{figure}

\begin{figure}
    \centering
    \includegraphics[width=1\columnwidth]{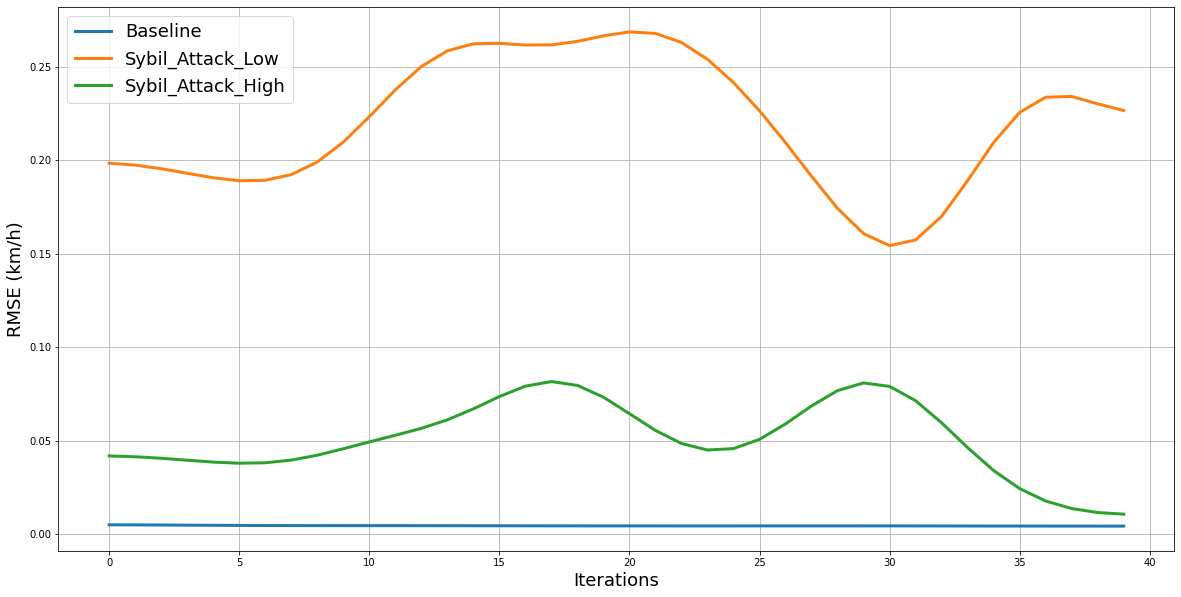}
    \caption{Impact of the Sybil\textunderscore Attack in different traffic densities.}
    \label{fig:Prediction_FL_Densities_2}
\end{figure}

We present in Figures~\ref{fig:Prediction_FL_Densities_1} and ~\ref{fig:Prediction_FL_Densities_2} the impact of the Single\textunderscore Attack and the Sybil\textunderscore Attack in terms of prediction accuracy of the LSTM model in a federated learning setting under different traffic densities. Under low traffic densities, the FL protocol takes into consideration all vehicles available on the link to participate and become \textit{workers} in the FL process. However, under higher densities, the FL protocol adapts and does not consider all the vehicles on the link as \textit{workers}. The protocol selects only a fixed amount of random vehicles to participate in the training in order to not congest the communication medium and address the optimal use of low bandwidth channels challenge. 

Consequently, under higher densities, the corrupted CAV of the Single\textunderscore Attack scenario may not be selected by the \textit{chief} and thus, as can be seen in Figure~\ref{fig:Prediction_FL_Densities_1}, the attack did not cause a big impact on the system. However, under low densities, the attack caused more damage on the performance of the global model. We notice an increase of approximately 0.115 in RMSE in low traffic density in comparison to an increase of only 0.025 under high densities.

On the other hand, the Sybil\textunderscore Attack has severe consequences in low densities and higher impact in high densities since it increases the probability of the malicious CAV to be chosen to participate in the FL process. As can be seen in Figure~\ref{fig:Prediction_FL_Densities_2}, the attack caused an increase of approximately 0.22 in RMSE in low traffic density and an increase of 0.045 in RMSE in high traffic density.

Poisoning resilience defense mechanisms are urgently required as FL in its standard form is susceptible to such adversarial attacks. The threat to security comes from the mobility of the vehicles and must be mitigated. A key element in V2X communication is the ability for vehicles and RSU to effectively and efficiently communicate. In fact, constant message exchange must be conducted in real time and since vehicles are constantly evaluating their environment and their position, real-time communication may be maliciously exploited. Methods to authenticate vehicles must be implemented in order to enable secure future V2X applications. Information from a CAV must be securely transmitted and compromised information sent out by a malicious vehicle must be immediately identified.

\section{CONCLUSION}

In this paper, we explored the vulnerability of FL in the vehicular networks, where CAVs can take advantage of their mobility, the wireless medium and the privacy that FL is designed to provide to corrupt the global training of a model. Our attacks demonstrate that FL in its standard form is vulnerable to mobile attackers exploiting the medium to perform model poisoning. Demonstrating robustness to attackers of the type considered in this paper is yet to be achieved. In future work, we plan to explore sophisticated defense strategies which can provide guarantees against the CAV attackers. In particular, encryption, localization, behavioral analysis and clustering may be promising detection mechanisms in this context. 

\section*{Acknowledgment}
This work was presented at the Workshop on Security Challenges in Intelligent Transportation Systems (SCITS) (WS01), IV2021.

\bibliographystyle{IEEEtran}
\bibliography{ref_CAV_FL}

\end{document}